\documentclass[aps,prx,reprint,superscriptaddress]{revtex4-1}

\usepackage{graphicx}
\usepackage[normalem]{ulem}
\usepackage{natbib}
\usepackage{amsmath,array,amssymb}
\usepackage{dcolumn}
\usepackage{bm}
\usepackage{hyperref}
\usepackage{color}
\usepackage{amsmath, amsthm, amssymb, amsfonts}
\usepackage{todonotes}

\newcommand{\ket}[1]{\vert#1\rangle}

\graphicspath{{img/}}

\begin{document}
	
	\title{A long-lived solid-state optical quantum memory for high-rate quantum repeaters}

	\author{Mohsen Falamarzi Askarani}
	\altaffiliation{These authors contributed equally to this work.}
	\affiliation{QuTech and Kavli Institute of Nanoscience, Delft University of Technology, 2628 CJ Delft, The Netherlands}
	
		\author{Antariksha Das$^*$}
	\affiliation{QuTech and Kavli Institute of Nanoscience, Delft University of Technology, 2628 CJ Delft, The Netherlands}
	
		\author{Jacob H. Davidson}
	\affiliation{QuTech and Kavli Institute of Nanoscience, Delft University of Technology, 2628 CJ Delft, The Netherlands}
	
	\author{Gustavo C. Amaral}
	\affiliation{QuTech and Kavli Institute of Nanoscience, Delft University of Technology, 2628 CJ Delft, The Netherlands}
	
	\author{Neil Sinclair}
	\affiliation{John A. Paulson School of Engineering and Applied Sciences, Harvard University, Cambridge, Massachusetts 02138, USA}
	\affiliation{Division of Physics, Mathematics and Astronomy, and Alliance for Quantum Technologies (AQT), California Institute of Technology, Pasadena, California 91125, USA}

	\author{Joshua A. Slater}
	\affiliation{QuTech and Kavli Institute of Nanoscience, Delft University of Technology, 2628 CJ Delft, The Netherlands}
	
		\author{Sara Marzban}
	\affiliation{QuTech and Kavli Institute of Nanoscience, Delft University of Technology, 2628 CJ Delft, The Netherlands}
	
	\author{Charles W. Thiel}
\affiliation{Department of Physics, Montana State University, Bozeman, MT, USA}

\author{Rufus L. Cone}
\affiliation{Department of Physics, Montana State University, Bozeman, MT, USA}

	\author{Daniel Oblak}
\affiliation{Institute for Quantum Science and Technology, and Department of Physics \& Astronomy, University of Calgary, Calgary, Alberta, T2N 1N4, Canada}

	\author{Wolfgang Tittel}
	\affiliation{QuTech and Kavli Institute of Nanoscience, Delft University of Technology, 2628 CJ Delft, The Netherlands}
	\date{\today}
	
\begin{abstract}
We argue that long optical storage times are required to establish entanglement at high rates over large distances using memory-based quantum repeaters. Triggered by this conclusion, we investigate the $^3$H$_6$ $\leftrightarrow$  $^3$H$_4$ transition at 795.325 nm of Tm:Y$_3$Ga$_5$O$_{12}$ (Tm:YGG). Most importantly, we show that the optical coherence time can reach 1.1 ms, and, using laser pulses, we demonstrate optical storage based on the atomic frequency comb protocol up to 100 $\mu$s as well as a memory decay time T$_M$ of 13.1 $\mu$s. Possibilities of how to narrow the gap between the measured value of T$_m$ and its maximum of 275 $\mu$s are discussed. In addition, we demonstrate quantum state storage using members of non-classical photon pairs. Our results show the potential of Tm:YGG for creating quantum memories with long optical storage times, and open the path to building extended quantum networks.
\end{abstract}

	
\maketitle

\emph{Introduction --}
The future quantum internet \cite{kimble2008quantum,WehnerQInternet} will enable one to share entanglement and hence quantum information over large distances -- ultimately between any two points on earth. To overcome attenuation in optical fibers and enable quantum communication in a scalable manner, quantum repeaters are needed \cite{OriginalQR,DLCZ,sangouard2011quantum,NeilPLRmultiplexing,Jiang}. 

Optical quantum memories \cite{lvovsky2009optical} are vital building blocks in many quantum repeater architectures \cite{Jiang}. They allow storing qubits, encoded into photons that have travelled over long distances, until feed-forward information becomes available. This information specifies which optical mode---including spectral and temporal modes---a stored qubit should occupy once it has been re-emitted from the memory. Note that the required mode assignment (also referred-to as mode mapping operation) can happen in a memory-internal manner, e.g. by controlling the moment at which a photon is re-emitted (aka read-out on demand) \cite{EIT_PRL, DLCZ, CRIB, AFC_spinwave}, or externally, e.g. by shifting the spectrum of the emitted photon  \cite{NeilPLRmultiplexing, GrimauSource}. 

To increase the rate at which a quantum repeater distributes entanglement, it is important that qubits can be added continuously to the memory -- not only once a previously stored qubit has been re-emitted but also while it is still being stored. Such multiplexed storage implies (a) the use of ensembles of absorbers and (b) that any memory-specific control operation, triggered by the absorption of a newly arriving qubit, must neither affect re-emission nor the possibility for mode mapping of a previously absorbed qubit. Stated differently, any control operation required after absorption of a qubit or a train of qubits must not introduce deadtime that prevents the memory from storing additional qubits. 

Unfortunately, the latter requirement of qubit independence (condition b) can be at odds with a high repetition rate. As we show below, one example is that of temporal multiplexing and read-out on demand in the so-called atomic frequency comb (AFC) quantum memory protocol, which requires one to temporarily map qubit states from optical coherence to spin coherence \cite{AFC2009}. This leads us to conclude that it is important to optimize the \emph{optical storage time}, i.e. the time during which qubits are stored as optical coherence, which can be excited using visible or near infrared light. Triggered by this conclusion, we investigate purely optical storage using thulium-doped yttrium gallium garnet (Tm$^{3+}$:Y$_3$Ga$_5$O$_{12}$ or Tm:YGG) -- a rare-earth-ion doped crystal (REIC) whose promising spectroscopic properties have been demonstrated previously \cite{PRBTmYGG,PRLTmYGG}. However, the potential of Tm:YGG for storing photonic qubits has not yet been established. Here we show that its optical coherence time T$_2$ can reach 1.1 ms, which is one of the longest times reported for any REIC \cite{1msT2ErYSO,6msT2Eu}. Motivated by this promising result, we investigate the prospects of Tm:YGG for optical quantum storage using the two-level AFC protocol, and demonstrate storage of laser pulses for up to 100 $\mu$s. This time is comparable with recent results obtained using Yb:Y$_2$SiO$_5$ \cite{50usEcho}, and exceeds all other reported results of storage of light in optical coherence with any REIC by at least a factor of 10. However, we also find that the memory decay time T$_m$ of around 13 $\mu$s is 20 times smaller than the T$_2$-imposed maximum of 275 $\mu$s, showing the presence of decoherence that does not affect the T$_2$ measurements. Before addressing possible reasons for this large gap, we confirm the possibility for feed-forward-based spectral mode mapping---an alternative to read-out on demand \cite{NeilPLRmultiplexing}---and we also show that quantum correlations between members of photon pairs persist throughout storage, i.e. that our memory can operate in the quantum regime. We conclude by briefly discussing how to improve memory efficiencies by means of cavity-enhanced light-matter interaction.

\begin{figure}[h]
	\centering
	\includegraphics[width=1\linewidth]{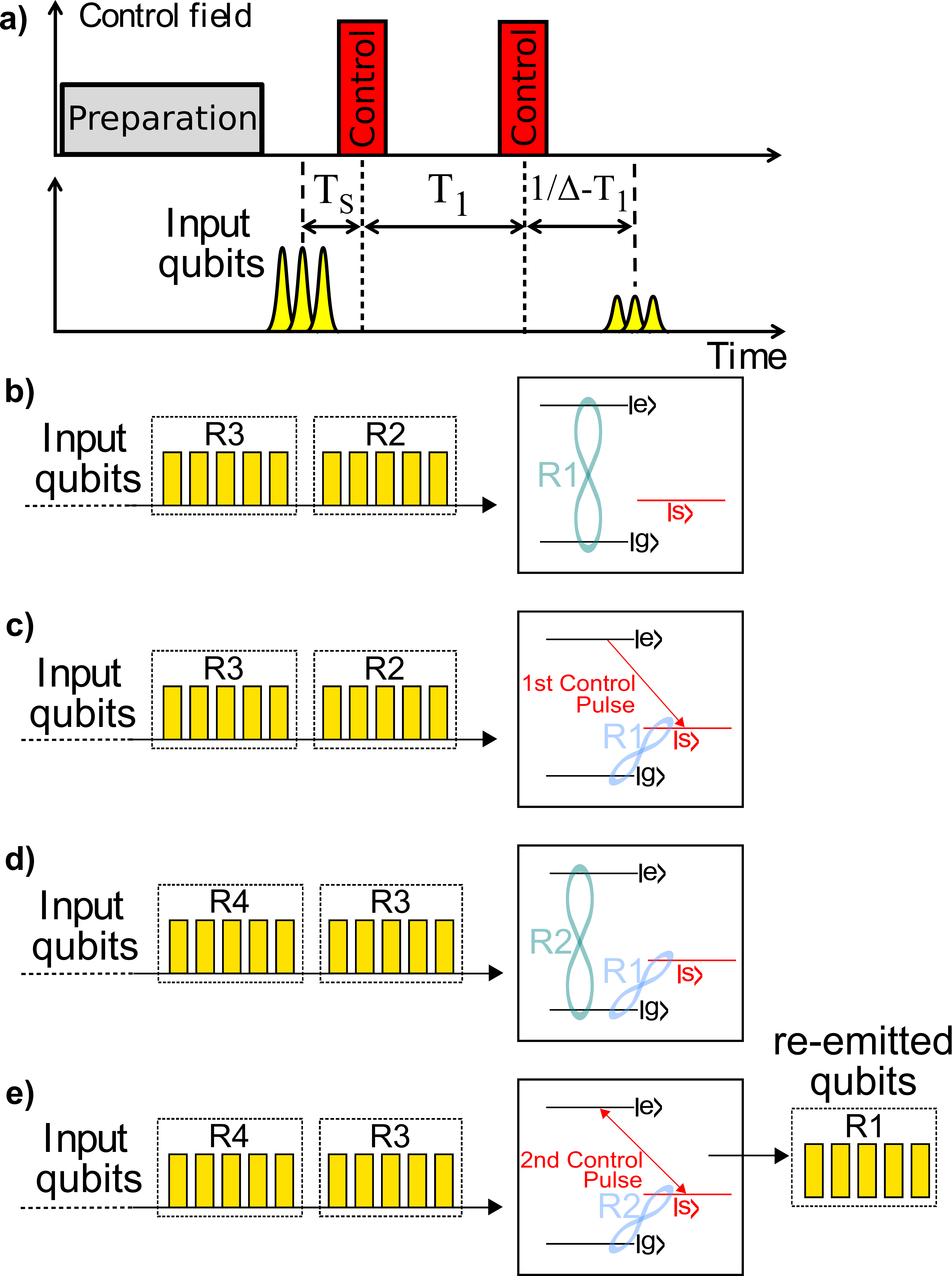}
	\caption{\textbf{Recall on demand using the AFC spin-wave protocol.} \textbf{a.} Pulse sequence for AFC spin-wave storage. A three-level lambda system is formed by spin states $\ket{g}$ and $\ket{s}$, and by excited state $\ket{e}$. \textbf{b - e.} Memory input and output, as well as atomic coherence (indicated by light blue and green figures of eight) for different moments during storage (see text for details). Trains of qubits are labelled R1-R4.}
	\label{fig:spin-wave AFC}
\end{figure}

\emph{The need for long optical storage} --
To support our claim that qubit independence can be at odds with a high repetition rate, let us discuss the example of REICs and temporal mode mapping using the AFC spin-wave memory \cite{AFC_spinwave}. 
As depicted in Fig. \ref{fig:spin-wave AFC}a, a pair of optical control pulses ($\pi$-pulses that resonantly couple the excited state $\ket{e}$ with a ground state $\ket{s}$) allows one to reversibly map optical coherence---excited through  the  absorption  of photonic  qubits---onto a spin transition. In this case, the timing of the second control pulse determines the moment at which the photons will be re-emitted from the memory, allowing for readout on demand.  

Let us now assume that a number of qubits---a first train---has already been absorbed by the memory, that the first control pulse has been applied, and that a second train of qubits has just been added to the memory (Figs. \ref{fig:spin-wave AFC}b-d depict this sequence). At this point, the first train (R1) is stored in terms of spin coherence, and the second (R2) in terms of optical coherence. Unfortunately, the subsequent control pulse, applied to the memory with the goal to map the second train onto spin coherence, simultaneously maps the first train back onto optical coherence. This causes re-emission of these qubits at a time that is determined by the need to transfer the second train, rather than by feed-forward information that specifies what to do with the first.

Fortunately, this problem can be avoided by storing only one train at a time. But in order to maximize the repetition rate of the repeater, this block, and hence the time during which qubits are stored in optical coherence, should be as long as possible -- ideally as long as the total storage time. Obviously, long optical storage times are also needed in all quantum memories that do not employ mapping between optical and spin coherence \cite{NeilPLRmultiplexing}.

\begin{figure}[tt]
	\centering
	\includegraphics[width=1\linewidth]{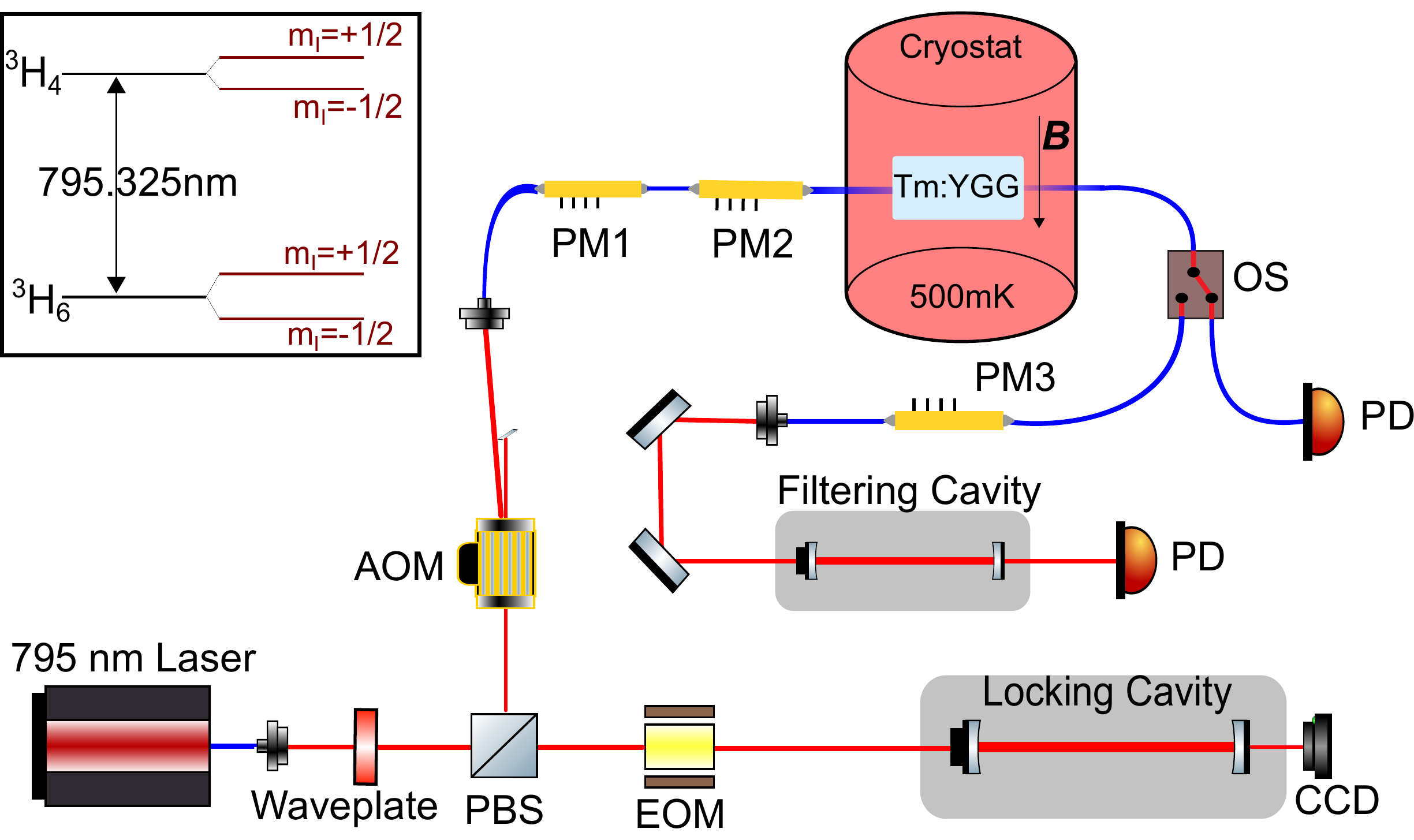}
	\caption{\textbf{Schematic of the experimental setup}. AOM: acousto-optic modulator; PM: phase modulator; OS: optical switch; EOM: electro-optic modulator; PD: photo detector; CCD: charge-coupled device camera; \textit{\textbf{B}}: Magnetic field. Inset: Simplified energy level diagram of Tm:YGG showing the $^3$H$_6\leftrightarrow ^3$H$_4$ zero-phonon line.}
	\label{fig:setup}
\end{figure}

\emph{Tm:YGG and experimental setup --} 
Due to their unique spectroscopic properties \cite{liu2006sREIs-book}, REICs have been broadly explored over the last two decades for quantum technology \cite{TheilREIDCsApplication,GoldnerREIDCsApplication}, including as ensemble-based quantum memory for light \cite{lvovsky2009optical,bussieres2013prospective, Heshimi_QuantumMemories} or for quantum processing \cite{saglamyurek2016multiplexed,kinos2021roadmap}. But while significant effort has been spent on achieving long storage times of qubits in spin coherence (values up to one hour have recently been reported \cite{onehourAFC,afzelius2014cavity,7.3usHugues_PRX,3.5usHugues_PRL}), much less work has been devoted to pushing and better understanding the limitations of storage in optical coherence. 

To address this shortcoming, we use a 25-mm long, 1\% Tm:YGG single crystal, mounted inside an adiabatic demagnetization refrigerator and cooled to around 500~mK. YGG forms a cubic lattice in which Tm$^{3+}$ replaces yttrium ions without charge compensation in six crystallographically equivalent sites of local D$_2$ point group symmetry \cite{D2_YGG}. A superconducting solenoid allows applying a magnetic field of up to 2 T along the crystal $<111>$ axis, resulting in the splitting of all electronic levels into two hyperfine sub-levels.
Fig. \ref{fig:setup} (Inset) depicts a simplified level structure.
 
The $^3$H$_6 \leftrightarrow  ^3$H$_4$ zero-phonon line at 795.325 nm is addressed by a tunable continuous-wave diode laser that propagates along the $<110> $ crystal direction. The polarization state of the input light is unknown but constant during the measurements, due to the use of a non-polarization maintaining fiber at the input of our crystal. The laser is frequency-locked to a high finesse cavity using the Pound-Drever-Hall (PDH) method, resulting in a frequency instability of less than 20 kHz over around 100 $\mu$s. To intensity- and frequency-modulate the light, we steer the beam into a single-pass acousto-optic modulator (AOM) (driven by a synthesized signal generator) and a phase modulator (PM) (driven by an arbitrary waveform generator). 
After passing through the crystal, the light is directed to a photo-detector. This setup is used for AFC creation (see \cite{GrimauPRR2QM} for a description), initial memory characterization, and storage of optical pulses in a single spectral mode. To allow for frequency-multiplexed storage and feed-forward recall, additional phase modulators are used to add frequency side-bands to the laser light, each of which creates a memory in a different spectral segment, and to frequency shift the light after re-emission so that only the desired spectral mode passes through an optical filter cavity \cite{NeilPLRmultiplexing}. See Fig. \ref{fig:setup} for a schematic.

\begin{figure*}[tttt]
	\centering
	\includegraphics[width=1\linewidth]{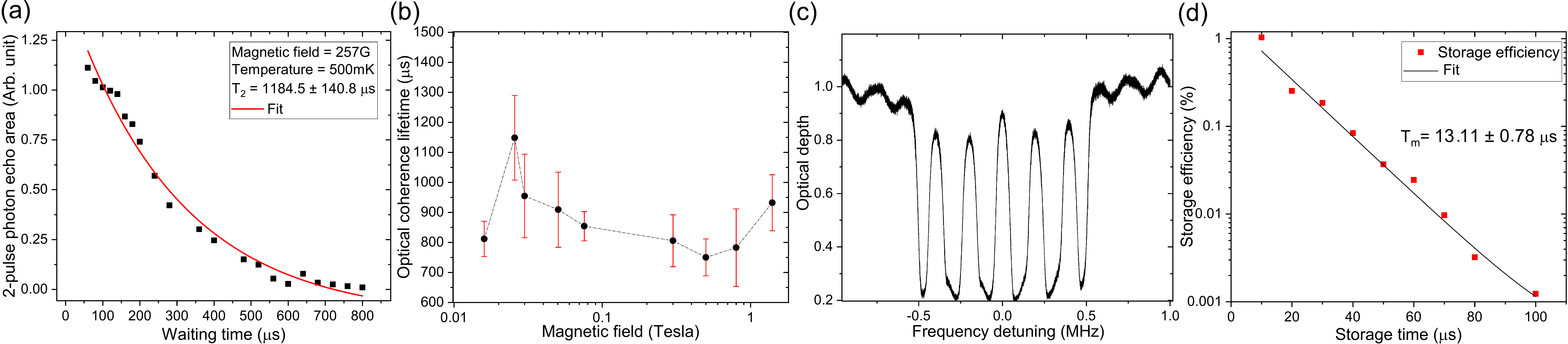}
	\caption{\textbf{Storage of data in a single spectral mode.} \textbf{a.} Exponential decay of the two-pulse photon echo signal at 257 G \textbf{b.} Optical coherence time T$_2$ as a function of magnetic field. The dashed line is a guide for the eye. \textbf{c.} AFC of 1 MHz bandwidth tailored for 5 $\mu$s storage time. \textbf{d.} Memory efficiency as a function of storage time using AFCs with finesse 2. The error bars in panels a and d are smaller than the data points.}
	\label{fig:memory1}
\end{figure*}

\emph{Measurements and results --} First, as a key property that determines the maximum optical storage time, we characterized the optical coherence time $T_2$ as a function of magnetic field using 2-pulse photon echoes \cite{TmLiNbO32PPE}. As an important difference compared to our previous studies \cite{PRBTmYGG, PRLTmYGG}, the temperature was lowered from 1.2 K to 500 mK. As shown in Figs. \ref{fig:memory1}a and \ref{fig:memory1}b and predicted by our earlier studies \cite{PRBTmYGG}, this resulted in a very significant improvement of the coherence time from 490 $\mu$s to around 1.1 ms, which makes the transition almost radiatively limited (the radiative lifetime T$_1$ of the $^3$H$_4$ level is 1.3 ms \cite{PRLTmYGG}) and corresponds to one of the longest reported optical coherence times for any rare-earth crystal.

Next, we investigated the possibility for optical data storage, both using laser pulses as well as quantum states of light. Towards this end, we employed the two-level atomic frequency comb protocol \cite{AFC2009}. An AFC is characterized by an absorption profile composed of evenly-spaced peaks in the frequency domain, which can be created using frequency-selective optical pumping of population from the troughs of the AFC to other atomic levels. Note that Tm:YGG is well suited for this task due to long-lived hyperfine levels within the electronic ground state manifold \cite{PRLTmYGG}. Absorption of a photon by an AFC results in the creation of a collective atomic excitation described by $\ket{\psi}_A=N^{-1/2}\sum_{j=1}^N c_j e^{i2\pi\delta_jt}e^{-kz_j}$, where $N$ is the number of atoms in the ensemble, $\delta_j$ the frequency detuning of the jth atom’s transition with respect to the input photon’s carrier frequency, and $z_j$ and $c_j$ denote the position and the excitation probability amplitude of the jth atom, respectively. After initial dephasing, the coherence rephases, resulting in re-emission of the photon after a time  $\tau$ that is determined by the inverse peak spacing $\Delta$, where $\tau=1/\Delta$. The AFC protocol is described in great detail in \cite{AFC2009,AFCMultimode}, and an example of an AFC in Tm:YGG is depicted in Fig. \ref{fig:memory1}c.

\emph{a) Long-lived storage of laser pulses --} Given the promisingly long optical coherence time, it is interesting to assess how the memory efficiency evolves as a function of storage time. To this end, we used laser pulses of 1 $\mu$s duration, and created AFCs with finesse F---the ratio between AFC peak spacing $\Delta$ and peak width $\delta$---of 2 (see \cite{NeilPLRmultiplexing,GrimauPRR2QM} for more information about AFC creation). Peak spacings varied between 100 kHz and 10 kHz, corresponding to storage times between 10 and 100 $\mu$s, respectively. The choice of F=2 maximizes the storage efficiency of the memory, which was limited due to the crystal's small optical depth of around 1. The AFC bandwidth for all storage times was 0.5~MHz except for 100 $\mu$s, where it was reduced to 0.2~MHz. The preparation of the AFCs took 1~s; it was followed by a waiting time of 20 ms to avoid detecting spurious photons caused by spontaneous decay of ions excited during the AFC preparation. The magnetic field in all measurements was around 100 G, which maximizes the optical coherence time. 

As shown in Fig. \ref{fig:memory1}d, we detected re-emitted pulses after up to 100 $\mu$s and found that the memory storage efficiency $\eta$ decreases exponentially as a function of storage time with a 1/e decay time of $T_m=$13.1$\pm$0.8 $\mu$s.
This value is much smaller than the ultimate limit imposed by T$_2$ of around 275 $\mu$s \cite{AFCMultimode}. This discrepancy shows that additional decoherence mechanisms are at play. Possibilities, some of which will be discussed in more detail in a forthcoming paper, include vibrations of the crystal inside the pulse-tube cooler, remaining laser frequency jitter, magnetic field instabilities, and spectral diffusion due to the interaction between thulium and neighboring ions in the crystal.

\begin{figure*}[ttt]
	\centering
	\includegraphics[width=0.85\linewidth]{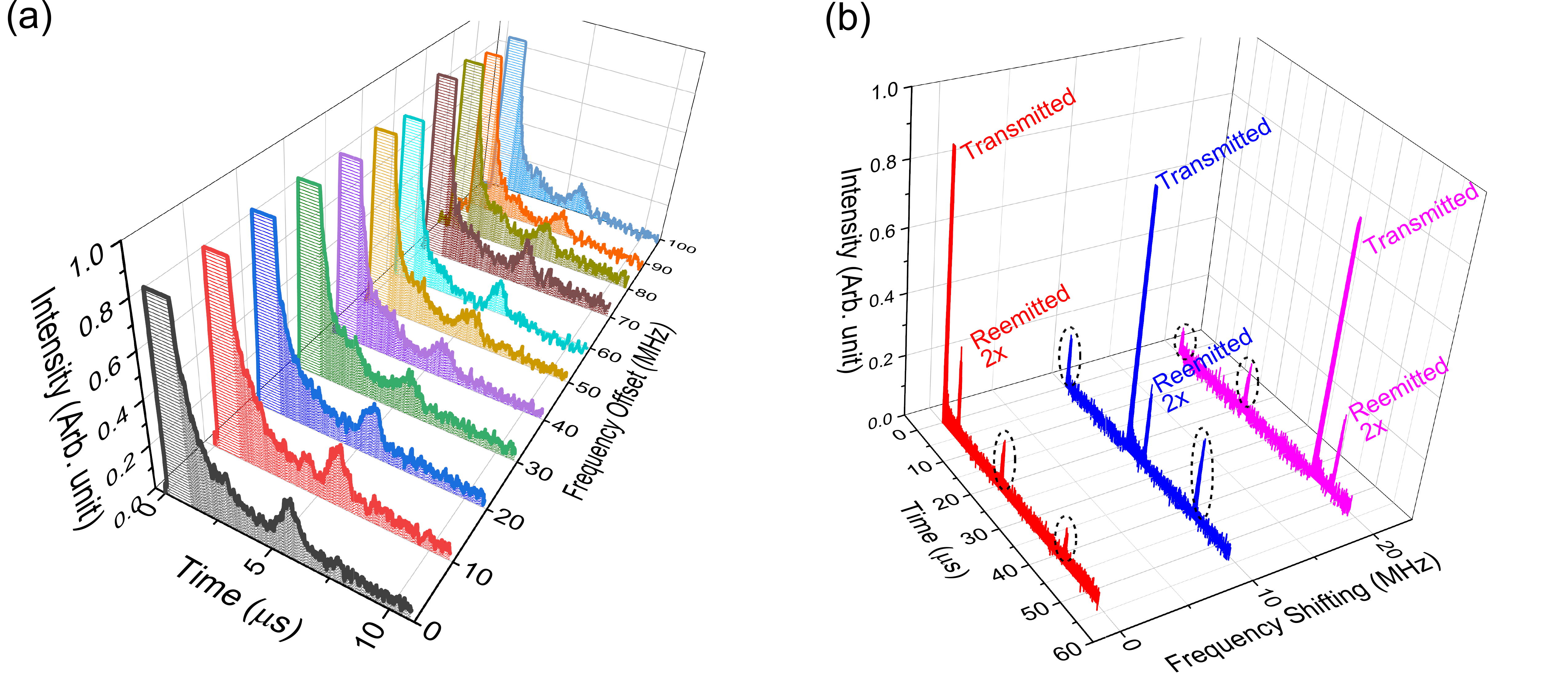}
	\caption{\textbf{Storage of data in multiple spectral modes.} \textbf{a.} Spectrally-multiplexed AFC quantum memory used to simultaneously store optical pulses in 11 spectral modes during 5 $\mu$s. The individual modes were resolved by changing the resonance frequency of the Fabry-Perot filtering cavity. \textbf{b.} Feed-forward mapping between different spectral modes. The resonance frequency of the filtering cavity only allowed transmission of the spectral mode at zero frequency detuning (depicted in red). The detuned frequency modes at 10 MHz and 20 MHz (shown in blue and pink, respectively) were resolved by shifting the frequency using a phase modulator (PM3). See description in the main text. The stored and re-emitted pulses are magnified by a factor of 2, and crosstalk is indicated using dotted circles.}
	\label{fig:memory2}
\end{figure*}

\emph{b) Frequency-multiplexed storage with feed-forward mode mapping -- } 
To demonstrate spectral multiplexing, we prepared 11 different AFCs (featuring F=2), each of 1 MHz bandwidth and spaced by 10 MHz, over a total bandwidth of 100 MHz.  Laser pulses of 1$\mu$s duration were created in each spectral mode. They were stored and recalled after 5$\mu$s (see Fig. \ref{fig:memory2}a). Assuming that five subsequent pulses fit into this time, this results in a multimode capacity over spectral and temporal degrees of freedom of 55. Note that the storage time---significantly shorter than our maximum of 100 $\mu$s in this and all subsequent measurements---was limited by a 
trade-off between more complex AFC tailoring procedures and memory efficiency. Otherwise, all parameters used to create the AFCs remained the same as before. 

For these measurements, the internal storage efficiency was close to 1.3\%. (This value was calculated by comparing the energies of input and re-emitted pulses and by taking coupling loss into account). Note that both the total bandwidth and the number of spectral channels can easily be increased with more laser power. Given the inhomogeneous broadening of the $^3$H$_6$$\leftrightarrow$$^3$H$_4$ transition of Tm:YGG of 56 GHz, this would in principle allow creating thousands of spectral modes. At the same time, our optimization of the comb-preparation parameters shows that more laser power would also improve the AFC quality per spectral mode and thereby allow increasing the storage time by an order of magnitude and hence the multi-mode capacity by another factor of ten. 

To meet the requirement of feed-forward-based mode mapping imposed by the targeted use of the memory in a quantum repeater, we furthermore demonstrated shifting of the recalled laser pulses such that only the desired spectral mode is subsequently transmitted through a filter cavity. For this proof-of-principle demonstration, which followed the approach introduced in \cite{NeilPLRmultiplexing}, we tailored with the help of PM2 three AFCs, again spaced by 10 MHz, in which spectrally matched optical pulses were stored. Pulses in different spectral modes were separated by 20~$\mu$s in order to make them distinguishable in time -- a necessary feature for analysis since the filter cavity after the memory erased all spectral distinguishability.

First, we set the resonance frequency of the filtering cavity such that it allowed transmission of the spectral mode at zero frequency detuning-- both for the light that was directly transmitted through the AFC as well as the stored and re-emitted pulse. The measured signal is depicted in Fig.~\ref{fig:memory2}b in red. At the same time, the cavity largely suppressed the pulses at 10 and 20 MHz detuning. However, because the cavity linewidth of 7.5 MHz is comparable with the spectral mode spacing of 10 MHz, small fractions of the neighboring modes leaked through. This crosstalk is visible in the peaks in the red-colored signal trace centered at 20 and 40 $\mu$s.

Next, we drove  PM3, positioned in-between the memory and the filter cavity, using a 10 MHz serrodyne signal that shifted all pulses emitted by the memory by -10 MHz. Consequently, the signal encoded originally at +10 MHz detuning, depicted in Fig.~\ref{fig:memory2}b in blue, became resonant with the cavity, resulting in its transmission through the cavity and the detection of a large peak at 20 $\mu$s followed by a smaller peak---the recalled pulse---5 $\mu$s later. As before, some leakage of signals in neighboring modes lead to peaks at 0 and 40 $\mu$s. 

Finally, the same experiment was repeated with a serrodyne shift of -20 MHz. The result is depicted by the magenta-colored signal trace in Fig.~\ref{fig:memory2}b.

\emph{c) Storage of heralded single photons --} Finally, we verified that Tm:YGG, in conjunction with the two-level AFC protocol, is suitable for quantum state storage. As described in detail in \cite{GrimauSource}, we created pairs of quantum-correlated photons at 795 and 1532 nm wavelength by means of spontaneous parametric down-conversion of strong laser pulses in a periodically-poled LiNbO$_3$ crystal. The detection of a 1532 nm photon using a superconducting nanowire single-photon detector (SNSPD) heralded the presence of a 795 nm photon, which was directed into, stored in, and released after 43 ns from the Tm:YGG memory. Note that the memory creation procedure remained unchanged except that the AFC bandwidth was increased to 4 GHz to better match the photon bandwidth, and that the magnetic field was increased to 12 kG to match the Tm level splitting with the spacing between a trough and the neighboring tooth in the AFC \cite{DavidsonCavity,saglamyurek2011broadband}. The photons were then detected using a single-photon detector based on a silicon avalanche photodiode. The storage efficiency, assessed by comparing photon detection rates with and without memory and after taking coupling loss into account, was around 0.35\%. 

To verify that the non-classical correlations with the 1532 nm photons persist throughout the storage process, we measured the 2nd order cross-correlation coefficient $g^{(2)}_{12}(t)$ between the two photons belonging to the same pair using time-resolved coincidence detection\cite{GrimauPRR2QM}. Before storage, we found $g ^{(2)}_{12}(0)_{no\:mem}$= 18$\pm$0.02 and, importantly, after storage $g^{(2)}_{12}(43)_{mem}$= 4.58$\pm$0.46. Both values surpass the classical upper bound of 2, confirming the quantum nature of both the photon source as well as of our memory.

 \emph{Outlook and Conclusion --} Our investigations have resulted in an optical coherence time $T_2$ up to 1.1 ms and optical storage times of up to 100 $\mu$s. However, they also revealed a memory decay time of 13.1 $\mu$s -- a factor of 20 less than the limit imposed by T$_2$. To further increase the memory performance, the most important task is to identify and remove the source of this excessive decoherence. Even though our laser is frequency stabilized and the 500 mK plate inside the cryostat on which the crystal is mounted features vibration isolation, it is very likely that remaining instabilities affect our measurements \cite{vibration_cryo}. Another possible source of spectral diffusion is that of ion-ion interaction inside the Tm:YGG crystal. Reducing the Tm doping concentration (which affects Tm-Tm interactions), e.g. by using a longer crystal with the same optical depth, or choosing a magnetic field direction for which the 795 nm transition becomes insensitive to first order to magnetic field fluctuations \cite{zefoz1, zefoz2, Faron_ZEFOZ, Afzelius_ZEFOZ} may help improving memory performance. 

In addition to the small memory decay time, our demonstration currently suffers from small recall efficiencies for all storage times. 
Selecting a specific polarization state at the input of the crystal to optimize the photon-ion interaction is expected to improve the observed memory efficiency. In addition, we note that embedding the crystal inside an impedance-matched cavity allows circumventing the effect of small optical depth provided AFCs with large finesse and small background absorption d$_0$ can be created  \cite{IMCavity2010,sabooni2013cavity,DavidsonCavity}. For instance, tailoring and analyzing an AFC with F = 4 allowed us to predict an efficiency of 15\% for 30 $\mu$s storage time, assuming the quality of the AFCs does not change when embedding the crystal inside the cavity. This would already allow using the memory for a proof-of-principle demonstration of an elementary quantum repeater link. Naturally, the creation of AFCs with further reduced background will result in a further increase of the efficiency. We anticipate that these improvements will make Tm:YGG suitable for efficient establishment of entanglement across extended quantum networks as a part of a quantum repeater.  

\section*{Acknowledgments}
The authors would like to thank M. Grimau Puigibert, T.~Chakraborty and O.~P.~Casas for help during various stages of the experiment, and M.~Afzelius for discussions. We acknowledge funding through the Netherlands Organization for Scientific Research (NWO), the European Union’s Horizon 2020 Research and Innovation Program under Grant Agreement No. 820445 and Project Name Quantum Internet Alliance, as well as by Alberta Innovates Technology Futures (AITF), the National Sciences and Engineering Research Council of Canada (NSERC) and the Alberta Ministry of Jobs, Economy and Innovation's Major Innovation Fund Project on Quantum Technologies. Furthermore, W.T. acknowledges funding as a Senior Fellow of the Canadian Institute for Advanced Research (CIFAR), and N.S. thanks the National Science Foundation Science and Technology Center for Integrated Quantum Materials under Cooperative Agreement No. DMR-1231319 as well as the AQT Intelligent Quantum Networks and Technologies (INQNET) research program for support. This material is based in part on research at Montana State University sponsored by Air Force Research Laboratory under agreement number FA8750-20-1-1004.\\

\bibliographystyle{apsrev4-1}
\bibliography{Tm-storage}
\end{document}